# Distributed Fuzzy Optimal Spectrum Sensing In Cognitive Radio

Dilip S. Aldar

**Abstract** –*The spectrum is a scarce resource and must utilize efficiently, the cognitive radio is a prospective solution for underutilized spectrum. The spectrum sensing is a key functionality to alleviate interference of secondary user to primary. The cognitive radios must detect the existence of the primary user and vacate the band for the primary immediately if the primary user detected. In cooperative sensing, every cognitive radio communicates their decision to fusion center via reporting channel. The reporting channels are not error free, which results corruption in the secondary's decision or information due to multipath fading and shadowing. This paper investigates the distributed fuzzy optimal cooperative spectrum sensing. The data and decision fusion with fuzzy detection is investigated in this paper. The simulation result shows the significant improvement in sensing performance over AND, OR and majority rules. The optimality in spectrum sensing is achieved by the proposed method with 1/3 of total malicious secondary users. The proposed scheme outperforms in the presence of malicious users*

## I. Introduction

The radio spectrum is one of the most scarce and precious resources due to exponential growth in wireless services. The allocated spectrum to the licensed or primary user (PU) is not fully utilized every time and every location, have been reported by spectral policy task force appointed by Federal communications commission [1]. Cognitive radio is a novel approach for improving the utilization of spectrum effectively [2] by potent techniques of spectrum sensing, spectrum management, and spectrum sharing.

The spectrum sensing must be performed efficiently by cognitive radio in the dynamically changing environment to detect the presence of primary user over a wide range of the spectrum. If the secondary user (SU) is in deep fade due to severe multipath fading may not detect the primary user and start using the spectrum already occupied by the primary, may cause the interference to primary user is referred hidden node problem [3]. To deal with the problems of propagation losses and interferences, cooperative spectrum sensing is one of best technique used. In the cooperative spectrum sensing approach every secondary user independently observe the presence or absence of primary by periodic sensing, and transmit the sensing information to the cognitive radio fusion center is referred information fusion or soft fusion which requires high bandwidth control channel, the control channel bandwidth should be at least same as the bandwidth of the sensed channel. The cognitive radio makes the decision and transmit to fusion center is referred decision fusion or hard fusion which requires low bandwidth control channel.

The reporting channels are not error free in practice, it is corrupted by white Gaussian noise, may results wrong reporting of the decision in hard fusion case, which is a serious issue and should be resolved before making the final decision. The energy detection method is widely used and most popular [4]. It is non-coherent with less implementation complexity and its performance is degraded in low SNR. The energy detection is utilized in this paper for cooperative spectrum sensing.

The hard decision cooperative spectrum sensing proposed in [5] to overcome the sensitivity requirement of individual radio and to analyze the effect of malfunctioning user in cooperative spectrum sensing. The cooperative cyclostationary techniques are proposed in [6] to improve the performance and reduce the complexity. Different approaches have been proposed for collaborative spectrum sensing in cognitive radio [7] includes collaborative wideband sensing, multiband joint detection, spatiospectral joint detection including energy detection, matched filtering and feature detection schemes whereas wavelet based centralized cooperative spectrum sensing proposed in [8] and SNR could be improved in low SNR case by using adaptive algorithms [9].

Analytical framework for cooperative spectrum sensing with data fusion was proposed in [4] and the performance of decision fusion was investigated with reporting error and without knowledge of primary signal SNR. The two step detector scheme proposed in [10] to deal with the noise through reporting channel from cognitive radio user to the cognitive base station.

In this paper, optimal fuzzy fusion scheme is proposed with the energy detector to improve the performance and reduce the complexity of cognitive radio systems. The cooperation among the cognitive users is analytically presented. The communication between the cognitive radio user and fusion center is not error free in practice, it is corrupted with the white Gaussian noise. The signal transmitted to the cognitive radio fusion center or cognitive radio base station must be detected correctly. The optimal fuzzy fusion scheme is presented for optimal detection and performance is evaluated in a different channel environment.

The rest of the paper is organized as follows. Section II discusses system model. Section III devoted to cooperative spectrum sensing over AWGN and multipath fading. Section IV design and analysis the Fuzzy based Fusion center, followed by the simulation results in section V. Section VI presents concluding remark.

## II. System Model

The secondary users need to sense the spectrum occupancy by the primary users. The signal received at each secondary user is modeled at $n^{th}$ time instant as binary hypothesis,

$H_0$: $r_i(n) = w_i(n)$, $i = 1,2,3 \ldots .,M$

$H_1$: $r_i(n) = h_i x_i(n) + w_i(n)$, $i = 1,2,3 \ldots .,M$, (1)

where $x_i(n)$ is the primary signal at $i^{th}$ secondary user, while $w_i(n)$ is the complex additive white Gaussian noise [11] with zero mean and variance $\sigma_i^2$, i.e., $w_i(n) \sim \mathcal{CN}(0, \sigma_i^2)$, Without loss of generality, $x_i(n)$ and $w_i(n)$ are assumed to be independent of each other. $h_i$ is the gain of the channel between the PU and the $i^{th}$ SU. $H_0$ denotes the PU is absent, and $H_1$ denotes the PU is present.

## III. Multiuser Cooperative Spectrum Sensing

In cooperative spectrum sensing, secondary users detect the band of interest of primary. The energy detection method is considered at secondary user to minimize the sensing overhead and is equipped with signal processing capabilities.

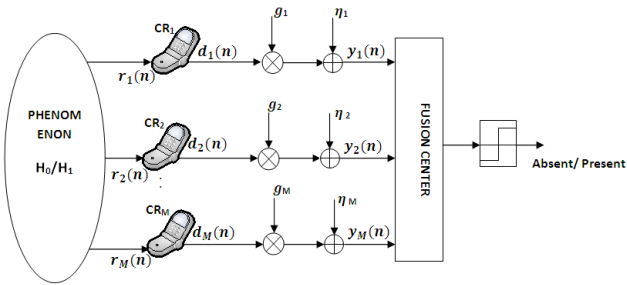

Fig.1. Cooperative Spectrum Sensing

Fig.1 shows the cooperative spectrum sensing cognitive radio system. The signal received by every secondary user is pre filtered by ideal bandpass filter has bandwidth W, and energy computed over the interval of N samples can be written as

$$s_i = \sum_{n=1}^{N} |r_i(n)|^2 , i = 1,2,3, \ldots \ldots, M, \quad (2)$$

where N=2WT, i.e. time bandwidth product.

If the signal is absent, the statistic $s_i$ follows the central chi-square distribution $\chi^2$ distribution with N degree of freedom, assuming that the uncertainty across secondary user are independent and identically distributed (i.i.d). If the signal is present the statistic $s_i$ follows the non central chi-square distribution $\chi^2$ distribution with N degree of freedom [12].

According to central limit theorem, the received energy signal $\{s_i\}_{i=1}^{M}$ are approximately normally distributed [13] with mean

$$E[s_i] = \begin{cases} N\sigma_i^2, & H_0 \\ N\sigma_i^2(1 + SNR), & H_1 \end{cases} \quad (3)$$

and variance

$$Var[s_i] = \begin{cases} 2N\sigma_i^4, & H_0 \\ 2N\sigma_i^4(1 + 2SNR), & H_1, \end{cases} \quad (4)$$

where SNR is signal to noise ratio = $SNR = \frac{|h_i|^2}{\sigma_i^2}$. The decision rule at each secondary user is given by

$$s_i \gtrless \gamma_i, \quad i = 1,2,3, \ldots \ldots, N \quad (5)$$

where $\gamma_i$ is decision threshold at each secondary user. The received statistics at each secondary user, is approximated as, $s_i \sim \mathcal{N}(E(s_i), Var(s_i))$, hence probability of false alarm is computed using [14],[15], are

$$P_f^{\ i} = P(s_i > \gamma_i | H_0) = Q\left(\frac{\gamma_i - N\sigma_i^2}{\sqrt{2N\sigma_i^4}}\right) \quad (6)$$

$$= Q\left(\frac{\gamma_i - N\sigma_i^2}{\sigma_i^2\sqrt{2N}}\right) \quad (7)$$

and,

$$P_d^{\ i} = P(s_i > \gamma_i | H_1) = Q\left(\frac{\gamma_i - N\sigma_i^2(1 + SNR)}{\sqrt{2N\sigma_i^4(1 + 2SNR)}}\right) \quad (8)$$

$$= Q\left(\frac{\gamma_i - N\sigma_i^2(1 + SNR)}{\sigma_i^2\sqrt{2N(1 + 2SNR)}}\right) \quad (9)$$

respectively.

The probability of the detection in terms of probability of the false alarm [15] is written as

$$P_d^{\ i} = Q\left(\frac{Q^{-1}(P_f^{\ i})\sigma_i^2\sqrt{2N} + N\sigma_i^2}{\sigma_i\sqrt{2N(1 + 2SNR)}}\right) \quad (10)$$

The false alarm and missed detection probabilities are tradeoffs between spectrum efficiency and interference to primary i.e., reliability. The lower the false alarm probability leads higher the spectrum utilization and larger the probability of detection (or lower the probability of miss detection, $P_m^{\ i} = 1 - P_d^{\ i}$) leads less interference to primary user.

## IV. Fuzzy Optimal Fusion Detection

This section presents the design of a fuzzy based fusion scheme for cognitive radio systems. The secondary users transmit the collected information about the primary over the reporting channel to fusion center is *information fusion*. The every secondary users performs the energy detection of the received primary signal and transmits the decision to the fusion center is *decision fusion*. In the cognitive radio system, the test statistics $\{s_i\}$ or decision at each secondary user $\{CR\}_1^M$, is transmitted to fusion center through reporting. The reporting channel is modeled as multipath faded channel, as illustrated in Fig.1. The received signal at fusion center is written as

$$y_i = y_i = g_i d_i + \eta_i, i = 1,2,\ldots\ldots, M \ , \qquad (11)$$

where $d_i = s_i$ denotes the test statistics is transmitted to the fusion center in the case of soft fusion and $d_i = 0|H_0$ or $d_i = 1|H_1$ when the decision is transmitted to the fusion center in the case of hard fusion. The $g_i$ are i.i.d. $\mathcal{CN}(0,\sigma_{gi}^2)$ multipath faded channel gains, $\eta_i$ is white Gaussian noise with zero mean and variance $\sigma_{\eta i}^2$, i.e., $\eta_i(n) \sim \mathcal{N}(0,\sigma_{\eta i}^2)$.

Three secondary users (i.e., M=3) are considered, which produces the crisp set of input data to the fuzzy logic system. The three antecedents are used to represent each secondary user, are characterized as linguistic variables. The received signal at the fusion center is $\{y_i\}$, therefore the input data are divided into three linguistic levels: LOW, MEDIUM, and HIGH. The consequent, i.e., the final decision has linguistic variable present and absent. The Membership functions in the fuzzification and defuzzification steps of a fuzzy logic system are shown in Fig.2. It is further used to map the non-fuzzy input values to fuzzy linguistic terms and vice versa. Triangular membership functions represent the antecedents and consequents are depicted in Fig. 2.

As M=3 therefore to setup the rule base for the fuzzy logic system, there are three antecedents and each antecedent having three fuzzy subsets. Therefore $3^3 = 27$ numbers of rule are formed as specified in table I. The Mamdani's fuzzy inference method [16] is used for the fuzzy inference process, to map the fusion center input to a decision output.

Defuzzification is the final step in the fuzzy logic system. There are different methods of defuzzification such as: centroid, bisector, Smallest of Maximum, and Middle of Maximum Largest of Maximum. All the methods of defuzzification are studied and investigated, but the most common method i.e., centroid defuzzification is chosen for the performance analysis of cooperative spectrum sensing.

## V. Simulation Results

Simulation results are presented in this section, different parameters considered to simulate the cognitive radio system are: bit rate 500 kb/s, maximum Doppler shift=200Hz, bits per frame = 200, carrier frequency of 2 GHz, delay vector = [0, 4, 8, 12] microseconds, gain Vector = [0, -3, -6, -9], Eb/N0=20 dB. Each secondary user has decision about the existence of primary or licensed user with the interference level $\{y_i\}$ in the range [-3 3], and energy statistics $\{s_i\}$ at secondary user in the range [0 150] is transmitted to fusion center are shown in Fig. 2.

TABLE I: Rules for Antecedent and Consequents

| Question# | Antecedent 1 | Antecedent 2 | Antecedent 3 | Consequent |
|---|---|---|---|---|
| 1 | LOW | LOW | LOW | ABSENT |
| 2 | LOW | LOW | MEDIUM | ABSENT |
| 3 | LOW | LOW | HIGH | ABSENT |
| 4 | LOW | MEDIUM | LOW | ABSENT |
| 5 | LOW | MEDIUM | MEDIUM | PRESENT |
| 6 | LOW | MEDIUM | HIGH | PRESENT |
| 7 | LOW | HIGH | LOW | ABSENT |
| 8 | LOW | HIGH | MEDIUM | PRESENT |
| 9 | LOW | HIGH | HIGH | PRESENT |
| 10 | MEDIUM | LOW | LOW | ABSENT |
| 11 | MEDIUM | LOW | MEDIUM | PRESENT |
| 12 | MEDIUM | LOW | HIGH | PRESENT |
| 13 | MEDIUM | MEDIUM | LOW | PRESENT |
| 14 | MEDIUM | MEDIUM | MEDIUM | PRESENT |
| 15 | MEDIUM | MEDIUM | HIGH | PRESENT |
| 16 | MEDIUM | HIGH | LOW | PRESENT |
| 17 | MEDIUM | HIGH | MEDIUM | PRESENT |
| 18 | MEDIUM | HIGH | HIGH | PRESENT |
| 19 | HIGH | LOW | LOW | ABSENT |
| 20 | HIGH | LOW | MEDIUM | PRESENT |
| 21 | HIGH | LOW | HIGH | PRESENT |
| 22 | HIGH | MEDIUM | LOW | PRESENT |
| 23 | HIGH | MEDIUM | MEDIUM | PRESENT |
| 24 | HIGH | MEDIUM | HIGH | PRESENT |
| 25 | HIGH | HIGH | LOW | PRESENT |
| 26 | HIGH | HIGH | MEDIUM | PRESENT |
| 27 | HIGH | HIGH | HIGH | PRESENT |

The performance of optimal fuzzy decision in information fusion is shown in Fig. 3. As shown in Fig.3 the energy statistics of three secondary users are CR1=56.9, CR2=82.2, CR3=85.8 respectively, and the optimal decision is observed at 0.695, for Information Fusion strategy. Similarly, Fig. 4 shows the performance of optimal fuzzy decision in decision fusion method. The decision about the primary user in the interested band by the secondary users are CR1=0.145, CR2=-0.506, CR3=-0.217. The optimal decision by decision fusion at fusion center is observed 0.695.

Fig.5 and Fig. 6 show the decision surface curve of the information fusion & decision fusion. The proposed method is also investigated in the presence of malicious and selfish secondary users. If the one third of the secondary users is malicious or selfish, this reports wrong decision and information to fusion center as shown in Fig.3 and Fig.4 respectively. The fuzzy optimal detection scheme gives correct decision about the primary user, irrespective of the decision and information of one third of the secondary users reported to the fusion center. Based on the decision or information received by the fusion center, on third of the malicious or selfish user could be punished or banned.

In order to analyze the detection performance of the proposed system majority rule is considered. Fig.7 shows the ROC curve of the Majority rule and optimal fuzzy detection, the majority rule is considered because it has reliable detection capability in multiuser scenario. It is observed that, the proposed scheme outperforms majority rule. The different fusion rules are also evaluated and observed that, OR rule always outperforms AND and majority rule, majority rule has better performance than AND rule.

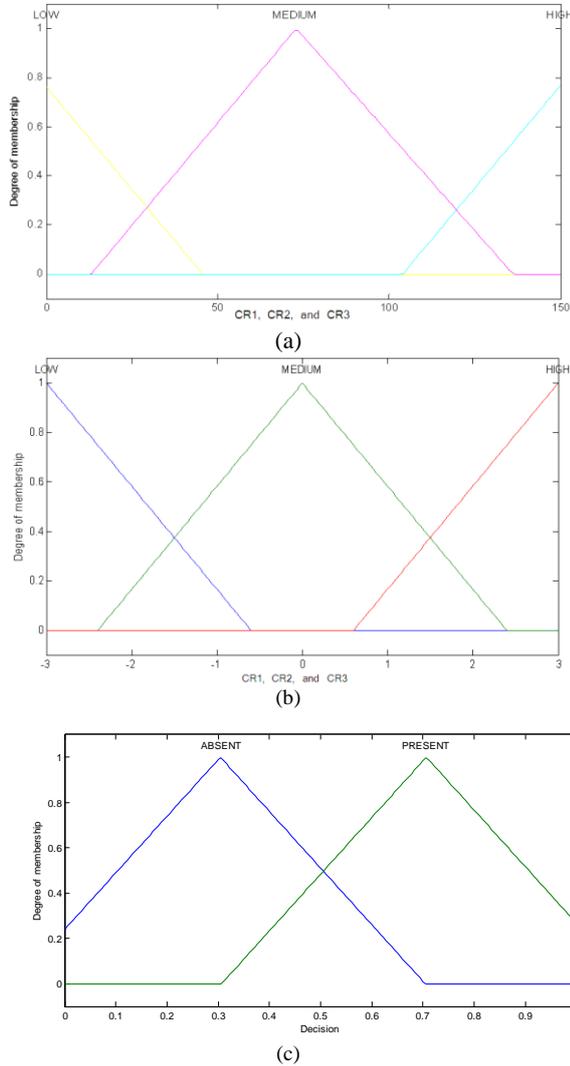

(a)

(b)

(c)

Fig. 2 .The membership functions used to represent: (a) and (b) Antecedent 1, Antecedent 2, Antecedent 3, and (c) Consequent, for information and decision fusion respectively.

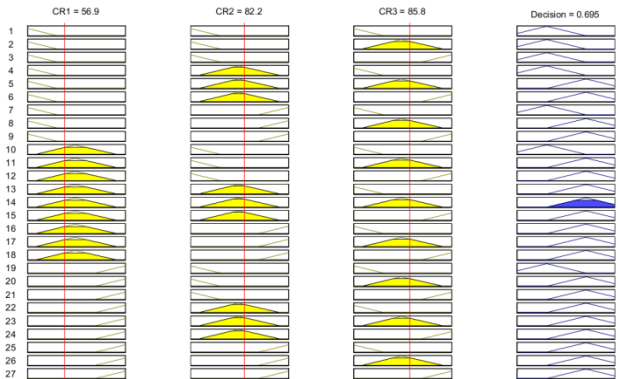

Fig. 3 Optimal Decision under difference secondary user combinations (CR1=56.9, CR2=82.2, CR3=85.8, Decision=0.695) for Information Fusion

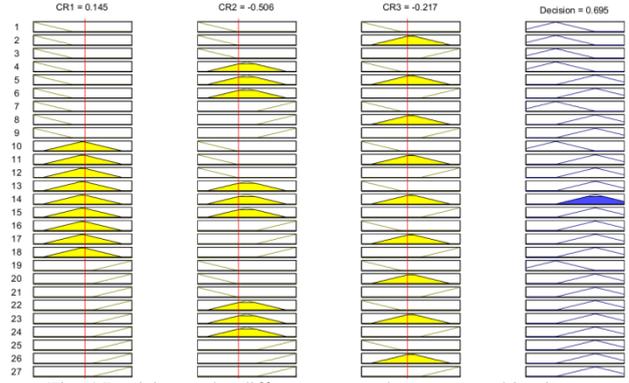

Fig. 4 Decision under difference secondary user combinations (CR1=0.145, CR2=-0.506, CR3=-0.217, Decision=0.695) for Decision fusion.

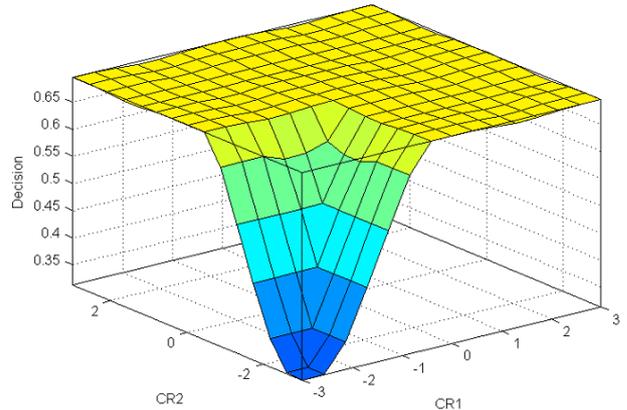

Fig. 5. Surface curve of fuzzy logic system for two secondary users in information fusion.

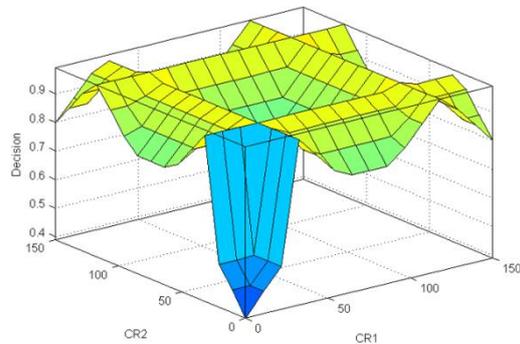

Fig. 6. Surface curve of fuzzy logic system for two secondary users in decision fusion.

The performance of a cognitive radio system is investigated in terms of probability of false alarm (Pf) and probability of detection (probability of miss detection Pm=1-Pd). Better spectrum efficiency is achieved by means of the fuzzy based spectrum sensing approach.The decision results after defuzzification of different methods are specified in table II. The Largest Maximum has better detection capability over all the other methods, the common centroid method used for the simulation, and the result of detection decision compared with majority rule as shown in Fig.7. The fuzzy optimal spectrum sensing scheme outperforms the majority rule.

## VI.  Conclusion

Spectrum sensing makes a pivotal contribution in cognitive radio system design, and efficient spectrum utilization. The cooperative spectrum sensing using fuzzy logic was proposed to make the optimal decision of available spectrum in the multipath fading environment and corrupted reporting channel.

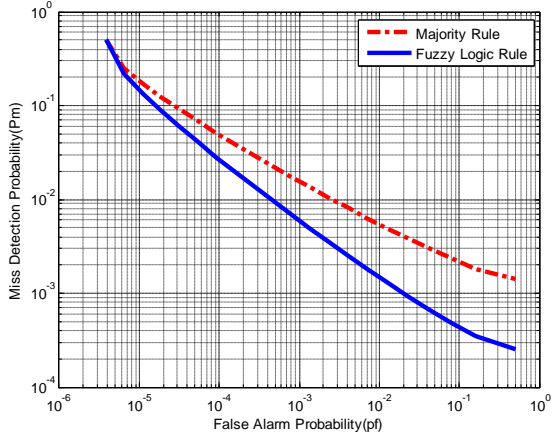

Fig.7. ROC curve to compare Majority fusion rule and Fuzzy fusion rule

TABLE II: THE CONSEQUENTS FOR THE DIFFERENT DEFUZZIFICATION METHODS

| Input to FLS | | Defuzzification Methods | Decision | |
|---|---|---|---|---|
| Information Fusion | Decision Fusion | | Information Fusion | Decision Fusion |
| CR1=78.61 CR2=60.54 CR3=76.81 | CR1=0.1446 CR2=-0.506 CR3=-0.2169 | Centroid | 0.695 | 0.695 |
| | | Bisector | 0.7 | 0.7 |
| | | Smallest of Maximum | 0.63 | 0.63 |
| | | Middle of Maximum | 0.705 | 0.71 |
| | | Largest of Maximum | 0.78 | 0.79 |

The Rayleigh multipath fading channel between primary user and secondary users, and AWGN channel for communication between secondary users and fusion center is considered. For the decision fusion strategy, OR, AND, Majority rules are studied for investigating the performance of the proposed system. The fuzzy logic fusion outperforms Majority rule, and the Majority rule has better detection capability over AND. In malicious or selfish user case, 1/3 of the total secondary users are the malicious users, their behavior did not affect on the decision performance and could be forced or punished for correct decision reporting.

The detection capability of proposed system also investigated in different defuzzification methods for information fusion and decision fusion, the Largest of Maximum method offers better decision compared to centroid method is considered here, which may results significant improvement in decision performance. The proposed approach offers optimum detection decision with corrupted energy statistics in information fusion and nosy decision in decision fusion strategies from secondary users. The proposed approach improves the spectrum efficiency significantly, and could be used for power control to improve power efficiency, and for power optimization in cognitive radio network.

## Authors' information

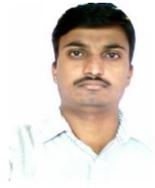

**Dilip S. Aldar** born in India received the B.Eng. degree from Shivaji University, India, in 2001, M.Tech from Dr. BATU, Lonere, India. He is currently working toward the Ph.D. degree in the area wireless communication.

He is currently Assistant professor in Electronics Engineering, K B P College of Engineering & polytechnic, Satara, India. He is working on Dynamic Spectrum Management and transmits power control algorithms for cognitive radio.

His area of research includes game theoretic approach for cognitive radio network, Genetic algorithms for cognitive radio.